\documentstyle[12pt]{article}

\def\lra{\leftrightarrow}
\def\Ph{{\hat P}}
\def\hs{\hspace{.1mm}}

\newcommand\as{\alpha_{\mathrm{S}}}

\newcommand\f[2]{\frac{#1}{#2}}
\def\ep{\epsilon}
\def\ktil{\widetilde k}

\def\beq{\begin{equation}}
\def\eeq{\end{equation}}
\def\beeq{\begin{eqnarray}}
\def\eeeq{\end{eqnarray}}
\def\cm{{\cal M}}

\def\to{\rightarrow}
\def\kper{k_{\perp}}

\newcommand{\la}{\langle}
\newcommand{\ra}{\rangle}

\def\nn{\nonumber}

\def\ID{1 \kern -.45 em 1}

\begin{document}

\begin{titlepage}
\renewcommand{\thefootnote}{\fnsymbol{footnote}}
\begin{flushright}
     CERN--TH/98-325\\ ETH--TH/98-27\\ hep-ph/9810389 
     \end{flushright}
\par \vspace{10mm}
\begin{center}
{\Large \bf
%Collinear-Factorization Formulae for\\[1ex] 
Collinear Factorization and Splitting Functions for\\[1ex] 
Next-to-next-to-leading Order QCD Calculations\footnote{This work was 
supported in part 
by the EU Fourth Framework Programme ``Training and Mobility of Researchers'', 
Network ``Quantum Chromodynamics and the Deep Structure of
Elementary Particles'', contract FMRX--CT98--0194 (DG 12 -- MIHT).}}
\end{center}
\par \vspace{2mm}
\begin{center}
{\bf Stefano Catani}~\footnote{On leave of absence from I.N.F.N.,
Sezione di Firenze, Florence, Italy.}\\

\vspace{5mm}

{Theory Division, CERN, CH 1211 Geneva 23, Switzerland} \\

\vspace{5mm}

{and}

\vspace{3mm}

{\bf Massimiliano Grazzini}\\

\vspace{5mm}
{Theory Division, CERN, CH 1211 Geneva 23, Switzerland} \\
and \\
Institute for Theoretical Physics, ETH-H\"onggerberg, CH 8093 Zurich, 
Switzerland~\footnote{Address after 1st October 1998.}

\vspace{5mm}

\end{center}

\par \vspace{2mm}
\begin{center} {\large \bf Abstract} \end{center}
\begin{quote}
\pretolerance 10000

We consider the singular behaviour of tree-level QCD amplitudes when the
momenta of three partons become simultaneously parallel.
We discuss the universal factorization formula that controls the singularities
of the multiparton matrix elements in this collinear limit and present the
explicit expressions of the corresponding splitting functions.
The results fully include spin (azimuthal) correlations
up to ${\cal O}(\as^2)$.
In the case of spin-averaged splitting functions we confirm
similar results obtained by Campbell and Glover.

\end{quote}

\vspace*{\fill}
\begin{flushleft}
     CERN--TH/98-325 \\ ETH--TH/98-27 \\  October 1998
\end{flushleft}
\end{titlepage}

\renewcommand{\thefootnote}{\fnsymbol{footnote}}
\section{Introduction}
\label{intro}

The universal properties of multiparton matrix elements in the infrared
(soft and collinear) limit play a relevant role in our capability 
to make reliable QCD predictions for hard-scattering 
processes [\ref{book}, \ref{lp97}]. 

At {\em leading} order in the QCD 
coupling $\as$, these properties are embodied in process-independent 
factorization formulae of tree-level [\ref{mangano}--\ref{antenna}] 
and one-loop
[\ref{BDKrev}--\ref{KST}, \ref{CSdipolelet}] amplitudes
that are at the basis of (at least) three important tools in perturbative QCD. 
The leading-logarithmic (LL) parton showers, which are implemented in Monte 
Carlo event generators [\ref{book}] to describe the exclusive structure of
hadronic final states, are based on these factorization
formulae supplemented with `jet calculus' techniques [\ref{KUV}] and 
colour-coherence properties [\ref{coher}, \ref{BCM}]. 
Analytical techniques
to perform all-order resummation of logarithmically enhanced contributions
at next-to-leading logarithmic (NLL) accuracy [\ref{softrev}] rely on the
factorization properties of soft and collinear emission.
More recently, the leading-order factorization formulae have been fully
exploited to set up completely general algorithms 
[\ref{GG}, \ref{GGK}--\ref{CSdipole}] to handle and cancel infrared 
singularities when
combining tree-level and one-loop contributions in the evaluation of jet cross
sections at the next-to-leading order (NLO) in perturbation theory.

Any further (although challenging) theoretical improvement of these tools 
requires the understanding of infrared-factorization properties at
the {\em next} order in $\as$. 

This topic has recently received considerable
attention. The infrared singular behaviour of two-loop QCD amplitudes has been 
discussed in Ref.~[\ref{sing2loop}].
The collinear limit of one-loop amplitudes is known [\ref{1loopcol}].
In the case of tree-level amplitudes, the 
factorization structure in the double-soft [\ref{dsoft}] 
and triple-collinear [\ref{glover}] limits has been studied.

In this letter, we reconsider the triple-collinear limit examined by
Campbell and Glover [\ref{glover}]. In particular, we extend their results
by fully taking into account azimuthal (spin) correlations. Besides
improving the general understanding of the collinear behaviour of
tree amplitudes, this extension is essential to apply some general
methods to perform exact fixed-order calculations at the 
next-to-next-to-leading order (NNLO) in perturbation theory. For instance,
the subtraction method [\ref{submeth}, \ref{CSdipole}] works by regularizing
the infrared singularities of the tree-level matrix element by identifying
and subtracting
a proper {\em local} counterterm. Thus, the knowledge of the
azimuthally {\em averaged} 
collinear limit studied in Ref.~[\ref{glover}]
is not sufficient for this
purpose.

The outline of the paper is as follows. In Sect.~\ref{kinem}, we define our
notation and the collinear kinematics and, after reviewing the known
collinear-factorization formulae at ${\cal O}(\as)$, we present their
generalization at the next perturbative order. Our explicit results
for the collinear-splitting functions at ${\cal O}(\as^2)$ are given in
Sect.~\ref{splitt}, where the comparison with the azimuthally-averaged case
considered in Ref.~[\ref{glover}] is also discussed. Details on our
calculation of the splitting functions are not discussed here and will 
be presented elsewhere [\ref{inprep}]. Some final remarks are left to 
Sect.~\ref{summa}.

\section{Notation and kinematics}
\label{kinem}

We consider a generic scattering process involving 
final-state\footnote{The case of incoming partons can be recovered by
simply crossing the parton flavours and momenta.}
QCD partons (massless quarks and gluons) with momenta 
$p_1, p_2, \dots$ Non-QCD partons $(\gamma^*, Z^0, W^\pm, \dots)$, carrying
a total momentum $Q$, are always understood. The corresponding tree-level
matrix element is denoted by
\beq
\label{meldef}
\cm^{c_1,c_2,\dots;s_1,s_2,\dots}_{a_1,a_2,\dots}(p_1,p_2,\dots) \;\;,
\eeq
where $\{c_1,c_2,\dots\}$, $\{s_1,s_2,\dots\}$ and $\{a_1,a_2,\dots\}$ are
respectively colour, spin and flavour indices. The matrix element squared
summed over final-state colours and spins will be denoted by 
$| \cm_{a_1,a_2,\dots}(p_1,p_2,\dots) |^2$. To the purpose of the present
paper,
it is also useful to consider the sum over the spins of all the final-state
partons but one. For instance, if the sum over the spin polarizations of 
the parton $a_1$ is not carried out, we define the following
`spin-polarization tensor'
\beq
\label{melspindef}
{\cal T}_{a_1,\dots}^{s_1 s'_1}(p_1,\dots) \equiv 
\sum_{{\rm spins} \,\neq s_1,s'_1} \, \sum_{{\rm colours}}
\cm^{c_1,c_2,\dots;s_1,s_2,\dots}_{a_1,a_2,\dots}(p_1,p_2,\dots) \,
\left[ \cm^{c_1,c_2,\dots;s'_1,s_2,\dots}_{a_1,a_2,\dots}(p_1,p_2,\dots)
\right]^\dagger
\;\;.
\eeq

In the evaluation of the matrix
element, we use dimensional regularization in $d=4 - 2\ep$ space-time
dimensions and consider two helicity states for massless quarks and 
$d-2$ helicity states for gluons. This defines the usual 
dimensional-regularization scheme. Thus, the fermion spin indices are $s=1,2$
while to label the gluon spin it is convenient to use the corresponding Lorentz
index $\mu=1, \dots, d$. The $d$-dimensional average of the matrix element
over the polarizations of a parton $a$ is obtained by means of the factors
\beq
\label{ferav}
\frac{1}{2} \delta_{ss'}
\eeq
for a fermion, and (the gauge terms are proportional either to $p^\mu$
or to $p^\nu$)
\beq
\label{gluav}
\frac{1}{d-2} d_{\mu \nu}(p) = \frac{1}{2(1-\ep)} ( - g_{\mu \nu} +
{\rm gauge \; terms} )
\eeq
with 
\beq
\label{dprop}
- g^{\mu \nu} d_{\mu \nu}(p) = d-2 \;, \;\;\;\;\;
p^\mu \,d_{\mu \nu}(p) = d_{\mu \nu}(p) \,p^\nu = 0 \;,
\eeq
for a gluon with on-shell momentum $p$.

The relevant collinear limit at ${\cal O}(\as)$ is approached when the momenta
of two partons, say $p_1$ and $p_2$, become parallel. Usually, this limit is 
precisely defined as follows:
\beeq
\label{clim}
&&p_1^\mu = z p^\mu + k_\perp^\mu - \frac{k_\perp^2}{z} 
\frac{n^\mu}{2 p\cdot n} \;\;, \;\;\; p_2^\mu =
(1-z) p^\mu - k_\perp^\mu - \frac{k_\perp^2}{1-z} \frac{n^\mu}{2 p\cdot n}\;\;,
\nonumber \\
&&s_{12} \equiv 2 p_1 \cdot p_2 = - \frac{k_\perp^2}{z(1-z)} \;\;,
\;\;\;\;\;\;\;\; k_\perp \to 0 \;\;.
\eeeq
In Eq.~(\ref{clim}) the light-like ($p^2=0$) vector $p^\mu$ denotes the
collinear direction, while $n^\mu$ is an auxiliary light-like vector, which 
is necessary to specify the transverse component $k_\perp$ ($k_\perp^2<0$)
($k_\perp \cdot p = k_\perp \cdot n = 0$) or, equivalently, how the collinear 
direction is approached.
In the small-$k_\perp$ limit (i.e.\ neglecting terms that are less singular 
than $1/k_\perp^2$), the square of the matrix element in Eq.~(\ref{meldef})
fulfils the following factorization formula [\ref{book}]
\beeq
\label{cfac}
| \cm_{a_1,a_2,\dots}(p_1,p_2,\dots) |^2 \simeq \frac{2}{s_{12}} \;
4 \pi \mu^{2\ep} \as 
\;{\cal T}_{a,\dots}^{s s'}(p,\dots) \;
{\hat P}_{a_1 a_2}^{s s'}(z,\kper;\ep) \;\;,
\eeeq
where $\mu$ is the dimensional-regularization scale.
The spin-polarization tensor ${\cal T}_{a,\dots}^{s s'}(p,\dots)$
is obtained by replacing the partons $a_1$ and $a_2$ on the right-hand side
of Eq.~(\ref{melspindef}) with a single parton denoted by $a$.
This parton carries the quantum numbers of the
pair $a_1+a_2$ in the collinear limit. In other words, its momentum is 
$p^\mu$ and its other quantum numbers (flavour, colour) are obtained according
to the following rule: anything~+~gluon gives anything  and 
quark~+~antiquark gives gluon.

The kernel ${\hat P}_{a_1 a_2}$ in Eq.~(\ref{cfac}) is the $d$-dimensional
Altarelli--Parisi splitting function [\ref{AP}]. 
It depends not only on the momentum
fraction $z$ involved in the collinear splitting $a \to a_1 + a_2$, but also on
the transverse momentum $\kper$ and on the helicity of the parton $a$ in the
matrix element $\cm_{a,\dots}^{c,\dots;s,\dots}(p,\dots)$.
More precisely, ${\hat P}_{a_1 a_2}$ is in general a matrix
acting on the spin indices $s,s'$ of the parton $a$ in the 
spin-polarization tensor ${\cal T}_{a,\dots}^{s s'}(p,\dots)$.
Because of these {\em spin correlations}, the spin-average square
of the matrix element $\cm_{a,\dots}^{c,\dots;s,\dots}(p,\dots)$
cannot be simply factorized on the right-hand side of Eq.~(\ref{cfac}).

The explicit expressions of ${\hat P}_{a_1 a_2}$,
for the splitting processes
\beq
\label{sppro}
a(p) \to a_1(zp + \kper + {\cal O}(\kper^2)) +
a_2((1-z) p - \kper + {\cal O}(\kper^2)) \;\;,
\eeq
depend on the flavour of the partons $a_1, a_2$ and are given by
\beeq
\label{hpqqep}
{\hat P}_{qg}^{s s'}(z,\kper;\ep) = {\hat P}_{{\bar q}g}^{s s'}(z,\kper;\ep)
= \delta_{ss'} \;C_F
\;\left[ \frac{1 + z^2}{1-z} - \ep (1-z) \right] \;\;,
\eeeq
\beeq
\label{hpqgep}
{\hat P}_{gq}^{s s'}(z,\kper;\ep) = {\hat P}_{g{\bar q}}^{s s'}(z,\kper;\ep)
= \delta_{ss'} \;C_F
\;\left[ \frac{1 + (1-z)^2}{z} - \ep z \right] \;\;,
\eeeq
\beeq
\label{hpgqep}
{\hat P}_{q{\bar q}}^{\mu \nu}(z,\kper;\ep) 
= {\hat P}_{{\bar q}q}^{\mu \nu}(z,\kper;\ep)
= T_R
\left[ - g^{\mu \nu} + 4 z(1-z) \frac{\kper^{\mu} \kper^{\nu}}{\kper^2}
\right] \;\;,
\eeeq
\beq
\label{hpggep}
{\hat P}_{gg}^{\mu \nu}(z,\kper;\ep) = 2C_A
\;\left[ - g^{\mu \nu} \left( \frac{z}{1-z} + \frac{1-z}{z} \right)
- 2 (1-\ep) z(1-z) \frac{\kper^{\mu} \kper^{\nu}}{\kper^2}
\right] \;\;,
\eeq
where the $SU(N_c)$ QCD colour factors are 
\beq
\label{colofac}
C_F = \frac{N_c^2 -1}{2N_c} \;, \;\;\; C_A = N_c \;, 
\;\;\; T_R = \frac{1}{2} \;,
\eeq
and the spin indices of the parent parton $a$ have been denoted by $s,s'$
if $a$ is a fermion and $\mu,\nu$ if $a$ is a gluon.

Note that when the parent parton is a fermion (cf. Eqs.~(\ref{hpqqep}) and
(\ref{hpqgep})) the splitting function is proportional to the unity matrix
in the spin indices. Thus, in the factorization formula (\ref{cfac}),
spin correlations are effective only in the case of the collinear splitting
of a gluon. Owing to the $\kper$-dependence of the gluon splitting
functions in Eqs.~(\ref{hpgqep}) and (\ref{hpggep}), these spin correlations
produce a non-trivial azimuthal dependence with respect to the directions
of the other momenta in the factorized matrix element.

Equations (\ref{hpqqep})--(\ref{hpggep}) lead to the more familiar form of the
$d$-dimensional splitting functions only after average over the polarizations
of the parton $a$. The $d$-dimensional average is obtained by means of the
factors in Eqs.~(\ref{ferav}) and (\ref{gluav}).
Denoting by $\la {\hat P}_{a_1 a_2} \ra$
the average of ${\hat P}_{a_1 a_2}$ over the polarizations of the parent 
parton $a$, we have:
\beeq
\label{avhpqq}
\la {\hat P}_{qg}(z;\ep) \ra \, = \, \la {\hat P}_{{\bar q}g}(z;\ep) \ra \,
= C_F\;\left[ \frac{1 + z^2}{1-z} - \ep (1-z) \right] \;\;,
\eeeq
\beeq
\label{avhpqg}
\la {\hat P}_{gq}(z;\ep) \ra \, = \, \la {\hat P}_{g{\bar q}}(z;\ep) \ra \,
= C_F \;\left[ \frac{1 + (1-z)^2}{z} - \ep z \right] \;\;,
\eeeq
\beeq
\label{avhpgq}
\la {\hat P}_{q{\bar q}}(z;\ep) \ra \, = \, \la {\hat P}_{{\bar q}q}(z;\ep) \ra
 \, = T_R \left[ 1 - \frac{2 z(1-z)}{1-\ep} \right] \;\;,
\eeeq
\beq
\label{avhpgg}
\la {\hat P}_{gg}(z;\ep) \ra \, = \, 2C_A
\;\left[ \frac{z}{1-z} + \frac{1-z}{z}
+ z(1-z) \right] \;\;.
\eeq

In the rest of the paper we are interested in the collinear limit at
${\cal O}(\as^2)$. In this case three parton momenta can simultaneously become
parallel. Denoting these momenta by $p_1, p_2$ and $p_3$, their most general
parametrization is
\beq
\label{kin3}
p_i^\mu = x_i p^\mu +k_{\perp i}^\mu - \frac{k_{\perp i}^2}{x_i} 
\frac{n^\mu}{2p \cdot n} \;, \;\;\;\;\;i=1,2,3 \;,
\eeq
where, as in Eq.~(\ref{clim}), the light-like vector $p^\mu$ denotes 
the collinear direction and the auxiliary light-like vector $n^\mu$
specifies how the collinear direction is approached 
$(k_{\perp i} \cdot p = k_{\perp i} \cdot n = 0)$. Note that no other
constraint (e.g. $\sum_i x_i = 1$ or $\sum_i k_{\perp i} = 0$) is imposed
on the longitudinal and transverse variables $x_i$ and $k_{\perp i}$. Thus, we
can easily consider any (asymmetric) collinear limit at once.

In the triple-collinear limit, the matrix element squared
$| \cm_{a_1,a_2,a_3,\dots}(p_1,p_2,p_3,\dots) |^2$ has the
singular behaviour
$| \cm_{a_1,a_2,a_3,\dots}(p_1,p_2,p_3,\dots) |^2 \sim 1/(s s')$, where
$s$ and $s'$ can be either two-particle $( s_{ij} = (p_i + p_j)^2 )$
or three-particle $( s_{123} = (p_1 + p_2 + p_3)^2 )$ sub-energies. More
precisely, it can be shown [\ref{glover}, \ref{inprep}] that the 
matrix element squared still fulfils a factorization formula analogous to
Eq.~(\ref{cfac}), namely
\beeq
\label{ccfac}
| \cm_{a_1,a_2,a_3,\dots}(p_1,p_2,p_3,\dots) |^2 \simeq
\frac{4}{s^2_{123}} \; (4 \pi \mu^{2\ep} \as)^2 
\;{\cal T}_{a,\dots}^{s s'}(p,\dots) \;
{\hat P}_{a_1 a_2 a_3}^{s s'}
%(z_i,{\ktil}_i;\ep) 
\;\;.
\eeeq
Likewise in Eq.~(\ref{cfac}), the spin-polarization tensor 
${\cal T}_{a,\dots}^{s s'}(p,\dots)$ is obtained by replacing the partons 
$a_1$, $a_2$ and $a_3$ with a single parent parton, whose flavour $a$ is 
determined (see Sect.~\ref{splitt}) by flavour conservation in the 
splitting process $a \to a_1+a_2+a_3$. 

The three-parton splitting functions ${\hat P}_{a_1 a_2 a_3}$ generalize 
the Altarelli--Parisi splitting functions in Eq.~(\ref{cfac}). The spin 
correlations produced by the collinear splitting are taken into account by
the splitting functions in a universal way, i.e. independently of the 
specific matrix element on the right-hand side of Eq.~(\ref{ccfac}).
Besides depending on the spin of the parent parton, the functions 
${\hat P}_{a_1 a_2 a_3}$ depend on the momenta $p_1, p_2, p_3$.
However, due to their invariance under longitudinal boosts along
the collinear direction, the splitting functions can depend in a non-trivial
way only on the sub-energy ratios $s_{ij}/s_{123}$ and on the following
longitudinal and transverse variables:
 \beeq
\label{zvar}
z_i &=& \frac{x_i}{\sum_{j=1}^3 \,x_j} \;\;, \\
\label{kvar}
{\ktil}_i^\mu &=& k_{\perp i}^\mu - \frac{x_i}{\sum_{k=1}^3 \,x_k} \;
\sum_{j=1}^3 k_{\perp j}^\mu \;\;,
\eeeq
which automatically satisfy the constraints
$\sum_{i=1}^3 z_i = 1$ and  $\sum_{i=1}^3 {\ktil}_i = 0$. To simplify the
explicit expressions of the splitting functions, 
we find it convenient to introduce also the variables
\beq
\label{tvar}
t_{ij,k}\equiv 2 \,\f{z_i s_{jk}-z_j s_{ik}}{z_i+z_j} +
\f{z_i-z_j}{z_i+z_j} \,s_{ij} \;\;.
\eeq
The results of our calculation are presented in the next section.

\section{Collinear splitting functions at ${\cal O}(\as^2)$}
\label{splitt}

To evaluate the three-parton splitting functions, we use power-counting
arguments and the universal factorization properties of collinear
singularities. The method [\ref{jetcalc}] consists in directly computing
{\em process-independent} Feynman subgraphs in a physical gauge.
Details on the method and on our calculation are given in Ref.~[\ref{inprep}].
In the following we present the complete results for the spin-dependent
splitting functions.

The list of (non-vanishing) splitting processes that we have to consider is
as follows:
\beeq
\label{qqqprime}
&& q\to {\bar q}^\prime_1 + q^\prime_2 + q_3 \;\;, 
\;\;({\bar q} \to {\bar q}^\prime_1 + q^\prime_2 + {\bar q}_3 ) \;\;, \\
\label{qqq}
&& q\to {\bar q}_1 + q_2 + q_3 \;\;, 
\;\;({\bar q} \to {\bar q}_1 + q_2 + {\bar q}_3 )\;\;, \\
\label{ggq}
&& q\to g_1 + g_2 + q_3 \;\;,
\;\;({\bar q} \to g_1 + g_2 + {\bar q}_3 ) \;\;, \\
\label{gqq}
&& g\to g_1 + q_2 + {\bar q}_3 \;\;, \\
\label{ggg}
&& g \to g_1 + g_2 + g_3 \;\;. 
\eeeq
The superscripts in $q^\prime, {\bar q}^\prime$ denote fermions with different
flavour with respect to $q, {\bar q}$. The splitting functions for the
processes in parenthesis in Eqs.~(\ref{qqqprime}) and (\ref{qqq}) can be simply
obtained by charge-conjugation invariance, i.e. 
$\Ph_{{\bar q}^\prime_1 q^\prime_2 {\bar q}_3} =
\Ph_{{\bar q}^\prime_1 q^\prime_2 q_3}$ and
$\Ph_{{\bar q}_1 q_2 {\bar q}_3} =
\Ph_{{\bar q}_1 q_2 q_3}$.
In summary, we have to compute five independent splitting functions.

In the case of the splitting processes that involve a fermion as parent parton
(see Eqs.~(\ref{qqqprime})--(\ref{ggq})), we find that spin correlations are
absent. We can thus write
\beq
\label{qaver}
\Ph^{ss'}_{{\bar q}^\prime_1 q^\prime_2 q_3} = \delta^{ss'} \,
\la \Ph_{{\bar q}^\prime_1 q^\prime_2 q_3} \ra \;\;,
\eeq
and likewise for $\Ph^{ss'}_{{\bar q}_1 q_2 q_3}$ and
$\Ph^{ss'}_{g_1 g_2 q_3}$. This feature is completely analogous to the 
${\cal O}(\as)$ case and follows from helicity conservation in the
the quark--gluon vector coupling. 

The spin-averaged splitting function for non-identical fermions in the final
state is
\beq
\label{qqqprimesf}
\la \Ph_{{\bar q}^\prime_1 q^\prime_2 q_3} \ra \, = \f{1}{2} \, 
C_F T_R \,\f{s_{123}}{s_{12}} \left[ - \f{t_{12,3}^2}{s_{12}s_{123}}
+\f{4z_3+(z_1-z_2)^2}{z_1+z_2} 
+ (1-2\ep) \left(z_1+z_2-\f{s_{12}}{s_{123}}\right)
\right] \;\;.
\eeq

The analogous splitting function in the case of final-state fermions with 
identical flavour can be written in terms of that in Eq.~(\ref{qqqprimesf}),
as follows
\beq
\label{qqqsf}
\la \Ph_{{\bar q}_1q_2q_3} \ra \, =
\left[ \la \Ph_{{\bar q}^\prime_1q^\prime_2q_3} \ra \, + \,(2\lra 3) \,\right]
+ \la \Ph^{({\rm id})}_{{\bar q}_1q_2q_3} \ra \;\;,
\eeq 
where
\beeq
\label{idensf}
\la \Ph^{({\rm id})}_{{\bar q}_1q_2q_3} \ra \,
&=& C_F \left( C_F-\f{1}{2} C_A \right)
 \Biggl\{ (1-\ep)\left( \f{2s_{23}}{s_{12}} - \ep \right)\nn\\
&+& \f{s_{123}}{s_{12}}\Biggl[\f{1+z_1^2}{1-z_2}-\f{2z_2}{1-z_3}
    -\ep\left(\f{(1-z_3)^2}{1-z_2}+1+z_1-\f{2z_2}{1-z_3}\right) 
- \ep^2(1-z_3)\Biggr] \nn\\
&-& \f{s_{123}^2}{s_{12}s_{13}}\f{z_1}{2}\left[\f{1+z_1^2}{(1-z_2)(1-z_3)}-\ep
    \left(1+2\f{1-z_2}{1-z_3}\right)
    -\ep^2\right] \Biggr\} + (2\lra 3) \;\;.
\eeeq

The splitting function of the remaining quark-decay subprocess can be 
decomposed according to the different colour coefficients:
\beq
\label{qggsf}
\la \Ph_{g_1 g_2 q_3} \ra \, =
C_F^2 \, \la \Ph_{g_1 g_2 q_3}^{({\rm ab})} \ra \,
+ \, C_F C_A \, \la \Ph_{g_1 g_2 q_3}^{({\rm nab})} \ra  \;\;,
\eeq
and the abelian and non-abelian contributions are
\beeq
\label{qggabsf}
\la \Ph_{g_1 g_2 q_3}^{({\rm ab})} \ra \, 
&=&\Biggl\{\f{s_{123}^2}{2s_{13}s_{23}}
z_3\left[\f{1+z_3^2}{z_1z_2}-\ep\f{z_1^2+z_2^2}{z_1z_2}-\ep(1+\ep)\right]\nn\\
&+&\f{s_{123}}{s_{13}}\Biggl[\f{z_3(1-z_1)+(1-z_2)^3}{z_1z_2}+\ep^2(1+z_3)
-\ep (z_1^2+z_1z_2+z_2^2)\f{1-z_2}{z_1z_2}\Biggr]\nn\\
&+&(1-\ep)\left[\ep-(1-\ep)\f{s_{23}}{s_{13}}\right]
\Biggr\}+(1\lra 2) \;\;,
\eeeq
\beeq
\label{qggnabsf}
\la \Ph_{g_1 g_2 q_3}^{({\rm nab})} \ra \,
&=&\Biggl\{(1-\ep)\left(\f{t_{12,3}^2}{4s_{12}^2}+\f{1}{4}
-\f{\ep}{2}\right)+\f{s_{123}^2}{2s_{12}s_{13}}
\Biggl[\f{(1-z_3)^2(1-\ep)+2z_3}{z_2}\nn\\
&+&\f{z_2^2(1-\ep)+2(1-z_2)}{1-z_3}\Biggr]
-\f{s_{123}^2}{4s_{13}s_{23}}z_3\Biggl[\f{(1-z_3)^2(1-\ep)+2z_3}{z_1z_2}
+\ep(1-\ep)\Biggr]\nn\\
&+&\f{s_{123}}{2s_{12}}\Biggl[(1-\ep)
\f{z_1(2-2z_1+z_1^2) - z_2(6 -6 z_2+ z_2^2)}{z_2(1-z_3)}
+2\ep\f{z_3(z_1-2z_2)-z_2}{z_2(1-z_3)}\Biggr]\nn\\
&+&\f{s_{123}}{2s_{13}}\Biggl[(1-\ep)\f{(1-z_2)^3
+z_3^2-z_2}{z_2(1-z_3)}
-\ep\left(\f{2(1-z_2)(z_2-z_3)}{z_2(1-z_3)}-z_1 + z_2\right)\nn\\
&-&\f{z_3(1-z_1)+(1-z_2)^3}{z_1z_2}
+\ep(1-z_2)\left(\f{z_1^2+z_2^2}{z_1z_2}-\ep\right)\Biggr]\Biggr\}
+(1\lra 2) \;\;.
\eeeq

In the case of collinear decays of a gluon (see Eqs.~(\ref{gqq}, \ref{ggg})),
spin correlations are highly non-trivial.

The colour-factor decomposition of the
splitting function for the decay into a $q{\bar q}$ pair plus a gluon is
\beq
\label{gqqsf}
\Ph^{\mu\nu}_{g_1 q_2 {\bar q}_3}  \, =
C_F T_R \, \Ph_{g_1 q_2 {\bar q}_3}^{\mu\nu \,({\rm ab})} \,
+ \, C_A T_R\, \Ph_{g_1 q_2 {\bar q}_3}^{\mu\nu \,({\rm nab})}  \;\;,
\eeq
where the abelian and non-abelian terms are given by
\beeq
\label{gqqabsf}
\Ph^{\mu\nu \,({\rm ab})}_{g_1q_2{\bar q}_3} &=&
%\Biggl\{
-g^{\mu\nu}\Biggl[ -2 
+ \f{2 s_{123} s_{23} + (1-\ep) (s_{123} - s_{23})^2}{s_{12}s_{13}}\Biggr]\nn\\
&+& \f{4s_{123}}{s_{12}s_{13}}\left({\ktil}_{3}^\mu
    {\ktil}_{2}^\nu+{\ktil}_{\hs 2}^\mu
    {\ktil}_{3}^\nu-(1-\ep){\ktil}_{\hs 1}^\mu 
    {\ktil}_{1}^\nu \right)
%    \Biggr\} 
\;\;,
\eeeq
\beeq
\label{gqqnabsf}
\Ph^{\mu\nu \,({\rm nab})}_{g_1q_2{\bar q}_3} &=& \f{1}{4}
\,\Biggl\{ \f{s_{123}}{s_{23}^2}
\Biggl[ g^{\mu\nu} \f{t_{23,1}^2}{s_{123}}-16\f{z_2^2z_3^2}{z_1(1-z_1)}
\left(\f{{\ktil}_2}{z_2}-\f{{\ktil}_3}{z_3}\right)^\mu
\left(\f{{\ktil}_2}{z_2}-\f{{\ktil}_3}{z_3}\right)^\nu \,\Biggr]\nn\\
&+& \f{s_{123}}{s_{12}s_{13}} \Biggl[ 2 s_{123} g^{\mu\nu}
- 4 ( {\ktil}_2^\mu {\ktil}_3^\nu + {\ktil}_3^\mu {\ktil}_2^\nu
- (1-\ep) {\ktil}_1^\mu {\ktil}_1^\nu ) \Biggr] \nn\\
&-& g^{\mu\nu} \Biggl[ - ( 1 -2 \ep) + 2\f{s_{123}}{s_{12}} 
\f{1-z_3}{z_1(1-z_1)} + 2\f{s_{123}}{s_{23}} 
\f{1-z_1 + 2 z_1^2}{z_1(1-z_1)}\Biggr]\nn\\
&+& \f{s_{123}}{s_{12}s_{23}} \Biggl[ - 2 s_{123} g^{\mu\nu}
\f{z_2(1-2z_1)}{z_1(1-z_1)} - 16 {\ktil}_3^\mu {\ktil}_3^\nu 
\f{z_2^2}{z_1(1-z_1)} 
+ 8(1-\ep) {\ktil}_2^\mu {\ktil}_2^\nu \nn\\
&+& 4 ({\ktil}_2^\mu {\ktil}_3^\nu  + {\ktil}_3^\mu {\ktil}_2^\nu )
%\left(\f{2 z_2 z_3-z_1(z_2 -z_3)}{z_1(1-z_1)}-\ep\right) 
\left(\f{2 z_2 (z_3-z_1)}{z_1(1-z_1)}+ (1-\ep) \right)
%+ 8(1-\ep) {\ktil}_2^\mu {\ktil}_2^\nu
\Biggr] \Biggr\} + \left( 2 \leftrightarrow 3 \right)  \;\;.
\eeeq

In the case of gluon decay into three collinear gluons we find
\beeq
\label{gggsf}
\Ph^{\mu\nu}_{g_1g_2g_3} &=& C_A^2 
\,\Biggl\{\f{(1-\ep)}{4s_{12}^2}
\Biggl[-g^{\mu\nu} t_{12,3}^2+16s_{123}\f{z_1^2z_2^2}{z_3(1-z_3)}
\left(\f{{\ktil}_2}{z_2}-\f{{\ktil}_1}{z_1}\right)^\mu
\left(\f{{\ktil}_2}{z_2}-\f{{\ktil}_1}{z_1}\right)^\nu \;\Biggr]\nn\\
&-& \f{3}{4}(1-\ep)g^{\mu\nu}+\f{s_{123}}{s_{12}}g^{\mu\nu}\f{1}{z_3}
    \Biggl[\f{2(1-z_3)+4z_3^2}{1-z_3}-\f{1-2z_3(1-z_3)}{z_1(1-z_1)}\Biggr]\nn\\
&+& \f{s_{123}(1-\ep)}{s_{12}s_{13}}\Biggl[2z_1\left({\ktil}^\mu_2
    {\ktil}^\nu_2\hs\f{1-2z_3}{z_3(1-z_3)}+
    {\ktil}^\mu_3{\ktil}^\nu_3\hs
    \f{1-2z_2}{z_2(1-z_2)}\right)\nn\\
&+& \f{s_{123}}{2(1-\ep)} g^{\mu\nu}
    \left(\f{4z_2z_3+2z_1(1-z_1)-1}{(1-z_2)(1-z_3)}
    - \f{1-2z_1(1-z_1)}{z_2z_3}\right)\nn\\
&+& \left({\ktil}_2^\mu{\ktil}_3^\nu
   +{\ktil}_3^\mu{\ktil}_2^\nu\right)
    \left(\f{2z_2(1-z_2)}{z_3(1-z_3)}-3\right)\Biggr]\Biggr\}
    + (5\mbox{ permutations}) \;\;.
\eeeq

The splitting functions in Eqs.~(\ref{gqqabsf})--(\ref{gggsf})
can be averaged over the spin polarizations of the parent
gluon according to Eq.~(\ref{gluav}):
\beq
\label{gsfav}
\la \Ph_{a_1 a_2 a_3} \ra \, \equiv \frac{1}{2 (1 - \ep)} \,d_{\mu \nu}(p)
\; \Ph^{\mu\nu}_{a_1 a_2 a_3} \;\;.
\eeq
Performing the average we obtain
\beeq
\label{gqqabsfav}
\la \Ph^{({\rm ab})}_{g_1q_2{\bar q}_3} \ra \,&=&
%\Biggl\{
-2-(1-\ep)s_{23}\left(\f{1}{s_{12}}+\f{1}{s_{13}}\right)
+ 2\f{s_{123}^2}{s_{12}s_{13}}\left(1+z_1^2-\f{z_1+2z_2 z_3}{1-\ep}\right) 
\nn\\
&-&\f{s_{123}}{s_{12}}\left(1+2z_1+\ep-2\f{z_1+z_2}{1-\ep}\right)
- \f{s_{123}}{s_{13}}\left(1+2z_1+\ep-2\f{z_1+z_3}{1-\ep}\right)
%\Biggr\} 
\;,
\eeeq
\beeq
\label{gqqnabsfav}
\la \Ph^{({\rm nab})}_{g_1q_2{\bar q}_3} \ra
\,&=&\Biggl\{-\f{t^2_{23,1}}{4s_{23}^2}
+\f{s_{123}^2}{2s_{13}s_{23}} z_3
\Biggl[\f{(1-z_1)^3-z_1^3}{z_1(1-z_1)}
-\f{2z_3\left(1-z_3 -2z_1z_2\right)}{(1-\ep)z_1(1-z_1)}\Biggr]\nn\\
&+&\f{s_{123}}{2s_{13}}(1-z_2)\Biggl[1
+\f{1}{z_1(1-z_1)}-\f{2z_2(1-z_2)}{(1-\ep)z_1(1-z_1)}\Biggr]\nn\\
&+&\f{s_{123}}{2s_{23}}\Biggl[\f{1+z_1^3}{z_1(1-z_1)}
+\f{z_1(z_3-z_2)^2-2z_2z_3(1+z_1)}
{(1-\ep)z_1(1-z_1)}\Biggr] \nn\\
&-&\f{1}{4}+\f{\ep}{2}
-\f{s_{123}^2}{2s_{12}s_{13}}\Biggl(1+z_1^2-\f{z_1+2z_2z_3}{1-\ep}
\Biggr) \Biggr\}
+ (2\lra  3) \;\;,
\eeeq
\beeq
\label{gggsfav}
\la \Ph_{g_1g_2g_3} \ra \,&=& C_A^2\Biggl\{\f{(1-\ep)}{4s_{12}^2}
t_{12,3}^2+\f{3}{4}(1-\ep)+\f{s_{123}}{s_{12}}\Biggl[4\f{z_1z_2-1}{1-z_3}
+\f{z_1z_2-2}{z_3}+\f{3}{2} +\f{5}{2}z_3\nn\\
&+&\f{\left(1-z_3(1-z_3)\right)^2}{z_3z_1(1-z_1)}\Biggr]
+\f{s_{123}^2}{s_{12}s_{13}}\Biggl[\f{z_1z_2(1-z_2)(1-2z_3)}{z_3(1-z_3)}
+z_2z_3 -2 +\f{z_1(1+2z_1)}{2}\nn\\
&+&\f{1+2z_1(1+z_1)}{2(1-z_2)(1-z_3)}
+\f{1-2z_1(1-z_1)}{2z_2z_3}\Biggr]\Biggr\}
+ (5\mbox{ permutations}) \;\;.
\eeeq

The ${\cal O}(\as^2)$-collinear behaviour of tree-level QCD matrix elements
has been independently examined by Campbell and Glover [\ref{glover}].
Their study differs in many respects from our analysis. Taken for granted the
universal factorization formula (\ref{ccfac}), they compute the three-parton
splitting functions by directly performing the collinear limit of the explicit
expressions of the $\gamma^* \to$~four- and five-parton squared matrix
elements. Moreover, they treat the colour structure in a different way and
consider the collinear limit of the colour-ordered sub-amplitudes
[\ref{mangano}]. Finally, they neglect spin correlations and present only
the explicit expressions of the polarization-averaged splitting functions. 

By properly taking into account the differences in the colour treatment,
we have compared our results with those of Ref.~[\ref{glover}] and found
complete agreement\footnote{Note that the published version of 
Ref.~[\ref{glover}] contains two misprints (in Eqs.~(5.8) and (5.19)) that have
been corrected in the archive version hep-ph/9710255 v3.} 
for the spin-averaged splitting functions. To be precise, the colour-connected
splitting functions $P_{a_1 a_2 a_3 \to a}$ of Ref.~[\ref{glover}] are related
to our spin-averaged splitting functions as follows:
\beeq
\label{CGcomp}
\la \Ph_{{\bar q}^\prime_1 q^\prime_2 q_3} \ra \, &=& \frac{s_{123}^2}{4} \;
  T_R C_F \, P^{{\rm non-ident.}}_{q_3 {\bar q}^\prime_1 q^\prime_2 \to q} \;\;, \nn \\
\la \Ph^{({\rm id})}_{{\bar q}_1q_2q_3} \ra \, &=& \frac{s_{123}^2}{4} \;
  C_F \left( C_F - \frac{1}{2} C_A \right) 
  P^{{\rm ident.}}_{q_3 {\bar q}_1 q_2 \to q} \;\;, \nn \\
\la \Ph_{g_1 g_2 q_3}^{({\rm ab})} \ra \, &=& \frac{s_{123}^2}{4} \;
  P_{q_3 {\tilde g}_1 {\tilde g}_2 \to q} \;\;,  \nn \\
\la \Ph_{g_1 g_2 q_3}^{({\rm nab})} \ra \, &=& \frac{s_{123}^2}{4} \;
\f{1}{2}\left(P_{q_3g_1g_2\to q}+P_{q_3g_2g_1\to q}-P_{q_3{\tilde
g}_1{\tilde g}_2\to q}\right) \;\;, \\
\la \Ph^{({\rm ab})}_{g_1q_2{\bar q}_3} \ra \, &=& \frac{s_{123}^2}{4} \;
  P_{q_2 {\tilde g}_1 {\bar q}_3 \to {\tilde g}} \;\;, \nn \\
\la \Ph^{({\rm nab})}_{g_1q_2{\bar q}_3} \ra \, &=& \frac{s_{123}^2}{4} \;
\frac{1}{2} \left( P_{g_1 {\bar q}_3 q_2 \to g} + P_{{\bar q}_3 q_2 g_1 \to g}
- P_{q_2 {\tilde g}_1 {\bar q}_3 \to {\tilde g}} \right) \;\;, \nn \\
\la \Ph_{g_1g_2g_3} \ra \, &=& \frac{s_{123}^2}{4} \;
  \left( \frac{C_A}{2} \right)^2 \;
  \left[ P_{g_1 g_2 g_3 \to g} + (5 \;\mbox{permutations}) \right] \;\;.\nn
\eeeq
Owing to the completely different methods used by the two groups,
this agreement can be regarded as an important cross-check of the calculations.

\section{Summary}
\label{summa}

We have considered the three-parton collinear limit of tree-level QCD
amplitudes. In this limit the singular behaviour of the matrix element squared
is given by the universal factorization formula (\ref{ccfac}) and is controlled
by process-independent splitting functions, which are analogous to the
Altarelli--Parisi splitting functions. In Sect.~\ref{splitt} we have presented
the explicit expressions of the splitting functions at ${\cal O}(\as^2)$,
taking fully into account spin correlations. 

These splitting functions are one of the necessary ingredients needed to extend
QCD predictions at higher perturbative orders. In particular, they are relevant
to perform analytic resummed calculations beyond NLL accuracy and to set up
general methods to compute jet cross sections at NNLO. The knowledge of the
collinear splitting functions, when combined with a consistent analysis of
soft-gluon coherence, can also give prospects of improving the logarithmic
accuracy of parton showers available at present for Monte Carlo event
generators.

\noindent {\bf Acknowledgements}. \\
\noindent We thank Nigel Glover for discussions.
One of us (M.G.) would like to thank the 
Fondazione `Angelo della Riccia' and INFN for financial support.

\section*{References}

% references
\def\ac#1#2#3{Acta Phys.\ Polon.\ #1 (19#3) #2}
\def\ap#1#2#3{Ann.\ Phys.\ (NY) #1 (19#3) #2}
\def\ar#1#2#3{Annu.\ Rev.\ Nucl.\ Part.\ Sci.\ #1 (19#3) #2}
\def\cpc#1#2#3{Computer Phys.\ Comm.\ #1 (19#3) #2}
\def\ib#1#2#3{ibid.\ #1 (19#3) #2}
\def\np#1#2#3{Nucl.\ Phys.\ B#1 (19#3) #2}
\def\pl#1#2#3{Phys.\ Lett.\ #1B (19#3) #2}
\def\pr#1#2#3{Phys.\ Rev.\ D #1 (19#3) #2}
\def\prep#1#2#3{Phys.\ Rep.\ #1 (19#3) #2}
\def\prl#1#2#3{Phys.\ Rev.\ Lett.\ #1 (19#3) #2}
\def\rmp#1#2#3{Rev.\ Mod.\ Phys.\ #1 (19#3) #2}
\def\sj#1#2#3{Sov.\ J.\ Nucl.\ Phys.\ #1 (19#3) #2}
\def\zp#1#2#3{Z.\ Phys.\ C#1 (19#3) #2}

\begin{enumerate}

\item \label{book}
R.K.\ Ellis, W.J.\ Stirling and B.R.\ Webber, {\it QCD and collider 
physics} (Cambridge University Press, Cambridge, 1996) and references therein.

\item \label{lp97}
S.\ Catani, hep-ph/9712442, in 
Proc. of the {\it XVIII International Symposium on Lepton-Photon
Interactions}, LP97, eds. A.\ De Roeck and A.\ Wagner (World Scientific,
Singapore, 1998), p.~147 and references therein.

\item \label{mangano}
M.L.\ Mangano and S.J.\ Parke, \prep{200}{301}{91}
and references therein.

\item \label{CSdipolelet}
S.\ Catani and M.H.\ Seymour, \pl{378}{287}{96}.

\item \label{antenna} 
D.A.\ Kosower, \pr{57}{5410}{98}. 
 
\item \label{BDKrev}
Z.\ Bern, L.\ Dixon and D.A.\ Kosower, \ar{46}{109}{96} and references therein.

\item \label{GG}
W.T. Giele and E.W.N. Glover, \pr{46}{1980}{92}.

\item \label{KST}
Z.\ Kunszt, A.\ Signer and Z. Tr\'ocs\'anyi, \np{420}{550}{94}.

\item \label{KUV}
K.\ Konishi, A.\ Ukawa and G.\ Veneziano, \np{157}{45}{79}.

\item \label{coher}
B.I.\ Ermolaev and V.S.\ Fadin, JETP Lett. 33 (1981) 269. 

\item \label{BCM}
A.\ Bassetto, M.\ Ciafaloni and G.\ Marchesini, \prep{100}{201}{83};
Yu.L.~Dokshitser, V.A.\ Khoze, A.H.\ Mueller and S.I. Troian,
{\it Basics of Perturbative QCD} (Editions Fronti\`eres, Gif-sur-Yvette, 1991)
and references therein.

\item \label{softrev}
G.\ Sterman, in Proc. {\it 10th Topical Workshop on Proton-Antiproton
Collider Physics}, eds. R.\ Raja and J.\ Yoh (AIP Press, New York, 1996),
p.~608; 
S.\ Catani, 
% hep-ph/9709503, published
in Proc. of the {\it 32nd Rencontres de Moriond: QCD and High-Energy
Hadronic Interactions}, ed. J. Tran Than Van (Editions Fronti\`eres, Paris,
1997), p.~331 and references therein.

\item \label{GGK}
W.T. Giele, E.W.N. Glover and D.A. Kosower, \np{403}{633}{93}.

\item \label{submeth}
Z.\ Kunszt and D.E.\ Soper, \pr{46}{192}{92};
S.\ Frixione, Z.\ Kunszt and A.\ Signer, \np{467}{399}{96};
Z. Nagy and Z. Tr\'ocs\'anyi, \np{486}{189}{97};
S.\ Frixione, \np{507}{295}{97}.

\item \label{CSdipole}
S.\ Catani and M.H.\ Seymour, \np{485}{291}{97}
(E ibid. B510 (1998) 503).

\item \label{sing2loop}
S.\ Catani, \pl{427}{161}{98}.

\item \label{1loopcol}
Z.\ Bern, G.\ Chalmers, L.\ Dixon and D.A.\ Kosower, \prl{72}{2134}{94};
Z.\ Bern, L.\ Dixon, D.C.\ Dunbar and D.A.\ Kosower, \np{425}{217}{94};
Z.\ Bern, L.\ Dixon and D.A.\ Kosower, \np{437}{259}{95};
Z.\ Bern and G.~Chalmers, \np{447}{465}{95}.

\item \label{dsoft}
F.A.\ Berends and W.T.\ Giele, \np{313}{595}{89};
S.\ Catani, in Proceedings of the Workshop on {\it New Techniques for
Calculating Higher Order QCD Corrections}, report ETH-TH/93-01, Zurich (1992). 

\item \label{glover}
J.M.\ Campbell and E.W.N. Glover, \np{527}{264}{98}. 
% (hep-ph/9710255 v3).

\item \label{inprep}
S.\ Catani and M.\ Grazzini, CERN preprint in preparation.

\item \label{AP}
G.\ Altarelli and G.\ Parisi, \np{126}{298}{77}.

\item \label{jetcalc}
J.\ Kalinowski, K.\ Konishi and T.R.\ Taylor, \np{181}{221}{81};
J.\ Kalinowski, K.\ Konishi, P.N.\ Scharbach and T.R.\ Taylor, 
\np{181}{253}{81}; J.F.\ Gunion, J.\ Kalinowski and L.\ Szymanowski,
\pr{32}{2303}{85}.

\end{enumerate}

\end{document}